\documentclass[fleqn,usenatbib]{mnras}

\usepackage{newtxtext,newtxmath}
\usepackage[T1]{fontenc}

\DeclareRobustCommand{\VAN}[3]{#2}
\let\VANthebibliography\thebibliography
\def\thebibliography{\DeclareRobustCommand{\VAN}[3]{##3}\VANthebibliography}

\usepackage{graphicx}	% Including figure files
\usepackage{amsmath}	% Advanced maths commands
\title[Stellar Abundances in Simulated Galaxies]{The Stellar Chemical Abundances of Simulated Massive Galaxies at $z=2$}

\author[Kim et al.]{
Jee-Ho Kim,$^{1}$\thanks{E-mail: jk3sxp@virginia.edu}
Sirio Belli,$^{2}$
Rainer Weinberger$^{3}$
\\
$^{1}$University of Virginia,
530 McCormick Rd., 
Charlottesville, VA 22904-4325, USA\\
$^{2}$Dipartimento di Fisica e Astronomia, 
Università di Bologna, 
via Gobetti 93/2, 40122 Bologna, Italy\\
$^{3}$Canadian Institute for Theoretical Astrophysics, 60 St. George Street, Toronto, ON M5S 3H8, Canada
}

\date{Accepted XXX. Received YYY; in original form ZZZ}

\pubyear{2022}

\begin{document}
\label{firstpage}
\pagerange{\pageref{firstpage}--\pageref{lastpage}}
\maketitle

\begin{abstract}
We analyze the stellar abundances of massive galaxies ($\log M_\ast/M_\odot>10.5$) at $z=2$ in the IllustrisTNG simulation with the goal of guiding the interpretation of current and future observations, particularly from JWST.
We find that the effective size, $R_e$, of galaxies strongly affects the abundance measurements: both [Mg/H] and [Fe/H] are anti-correlated with $R_e$, while the relative abundance [Mg/Fe] slightly increases with $R_e$. The $\alpha$ enhancement as tracked by [Mg/Fe] traces the formation timescale of a galaxy weakly, and mostly depends on $R_e$. Aperture effects are important: measuring the stellar abundances within 1~kpc instead of within $R_e$ can make a large difference. These results are all due to a nearly universal, steeply declining stellar abundance profile that does not scale with galaxy size --- small galaxies appear metal-rich because their stars live in the inner part of the profile where abundances are high. The slope of this profile is mostly set by the gas-phase abundance profile and not substantially modified by stellar age gradients. The gas-phase abundance profile, in turn, is determined by the strong radial dependence of the gas fraction and star formation efficiency. We develop a simple model to describe the chemical enrichment, in which each radial bin of a galaxy is treated as an independent closed-box system. This model reproduces the gas-phase abundance profile of simulated galaxies, but not the detailed distribution of their stellar abundances, for which gas and/or metal transport are likely needed.
\end{abstract}


%%%%%%%%%%%%%%%%%%%%%%%%%%%%%%%%%%%%%%%%%%%%%%%%%%

%%%%%%%%%%%%%%%%% BODY OF PAPER %%%%%%%%%%%%%%%%%%

\section{Introduction} \label{sec:intro}

One of the most important properties of galaxies is their chemical composition, which can be measured independently for gas and stars. While the \emph{gas-phase metallicity} is sensitive to processes such as star formation, inflows, and outflows \citep[e.g.,][]{2019Maiolino}, the \emph{stellar metallicity} probes the past history of a galaxy, since stars inherit the metallicity of the gas from which they form. For example, the stellar metallicity of quiescent galaxies can be used to investigate their quenching, a process that happened several Gyrs earlier \citep{2015Peng}. Thus, observational and theoretical studies of stellar metallicity can provide crucial insights on the formation and evolution of galaxies.

Observations in the local universe have shown that more massive galaxies tend to have a higher metal content; this is true whether one considers gas-phase metallicity \citep{2004Tremonti} or stellar metallicity \citep{2005Gallazzi}. Multiple theories have been developed to explain this mass-metallicity relation \citep[see][and references therein]{2019Maiolino}. In a popular scenario, first proposed by \citet{1974Larson}, massive galaxies are more metal-rich because their deep potential wells are able to retain most of the metals ejected by supernova-driven winds, while in low-mass systems these winds are more effective at removing metals. Other scenarios include additional important ingredients such as gas inflows and star formation efficiency \citep[e.g.,][]{2007Dalcanton}.

More recently, it has become possible to study chemical compositions using cosmological simulations of galaxy formation \citep{2008Finlator, 2015DeRossi, 2017Bahe, 2019Trayford, 2021Hemler}, which treat processes self-consistently --- for example, star formation and
cosmological environment activity including galaxy mergers, gas inflows, chemical enrichment, gas flows induced by stellar feedback, and transport of different chemical elements. These simulations have been used to understand the emergence of gas-phase scaling relations \citep{2019Torrey, 2021vanLoon, 2021Henry} and the properties of local galaxies \citep{2018Christensen, 2018Grand, 2019Grand}, with some studies also discussing the stellar metal abundances in galaxy populations \citep{2016Guidi, 2016Ma, 2016Taylor}.

In addition to measuring the overall metal content, analyzing the abundance of individual elements can provide further constraints on the formation and evolution of galaxies. In particular, $\alpha$ elements such as oxygen and magnesium are mostly produced by core-collapse supernovae, which occur soon after the formation of short-lived massive stars ($M_* > 8 M_\odot$); while iron peak elements like iron and nickel are produced by type Ia supernovae, which occur up to several Gyr after the onset of star formation, as they form from lower mass stars with longer life spans.
Thus, galaxies that formed rapidly must feature large amounts of $\alpha$ elements in their stars compared to iron-peak elements. Measurements of the relative abundance $[\alpha/\text{Fe}] = \log(N_\alpha / N_\text{Fe}) - \log(N_\alpha / N_\text{Fe})_{\odot}$ in the local universe revealed that massive galaxies are more $\alpha$-enhanced, suggesting that they formed more rapidly than lower-mass galaxies \citep{2000Trager, 2010Thomas}. Theoretical studies indicate a fundamental plane of stellar mass, star formation rate, and $\alpha$ enhancement in star-forming galaxies \citep{2018Matthee}.

Studies of local galaxies are ultimately limited by the difficulty of distinguishing between the stars that formed \textit{in-situ} and those that were accreted from satellite galaxies. In order to disentangle this effect, it is necessary to directly observe the progenitors of massive galaxies at high redshift. However, obtaining the high-quality spectra that are required to measure stellar abundances is particularly challenging for distant galaxies. Measurements of $\alpha$ enhancement for representative samples have only recently become available at intermediate redshift \citep[$z \sim 0.7$;][]{2021Beverage, 2022Borghi}, while at $z \sim 1-2$ only a few individual measurements exist \citep{2016Kriek, 2019Kriek, 2020Jafariyazani, 2020Lonoce}. This observational landscape is expected to change substantially thanks to the spectroscopic capabilities of JWST, which will open up a new era for the study of stellar abundances and their evolution.

The interpretation of metal abundance observations is further complicated by the presence of radial gradients in galaxies. In the local universe, metallicity declines with galactocentric distance \citep[e.g.,][]{2015GonzalezDelgado}, while [$\alpha$/Fe] may feature a weak increase with radius \citep[e.g.,][]{2017VanDokkum}. The meaning of galaxy-integrated abundance measurements, therefore, can be different for galaxies with different gradients, sizes, or morphologies \citep{2022Zibetti}.

The main goal of this work is to use numerical simulations to guide the interpretation of stellar metallicity observations, particularly for galaxies at the peak of cosmic star formation history, at $z\sim2$. We aim to connect \emph{spatially resolved} abundance gradients, which are closely related to the physical processes shaping the formation and evolution of galaxies, to \emph{galaxy-wide} measurements, which are easier to obtain for distant galaxies. We focus on massive systems, which are key to understanding the assembly and formation of massive halos in the early universe.

The paper is laid out as follows: Section~\ref{sec:data} describes the TNG numerical simulation and the data used in the analysis; Section~\ref{sec:global_measurements} explores galaxy-wide stellar abundances, while Section~\ref{sec:profiles} focuses on the spatially resolved abundance distributions and the connection between stellar and gas metallicity. We discuss our results in Section~\ref{sec:discussion} and give a summary in Section~\ref{sec:conclusion}. 

\section{Data} \label{sec:data}

\subsection{Simulation} \label{subsec:simulation}
We analyze predictions of stellar metallicity from the IllustrisTNG (TNG) project \citep{TNG1, TNG2, TNG3, TNG4, TNG5}. The TNG project is the next generation simulation suite after the Illustris simulation \citep{2014Vogelsberger, 2014bVogelsberger, 2014Genel}, using updated physics and modelling methods to expand our understanding of galaxy formation processes. TNG incorporates gravity, ideal magnetohydrodynamics, radiative cooling, star and massive black hole formation, as well as feedback from stars and active galactic nuclei. The detailed implementation is discussed in \citet{2017Weinberger} and \citet{2018Pillepich}. The simulation uses a flat $\Lambda$CDM cosmological model with $h = 0.6774$, $\Omega_{\Lambda,0} = 0.6911$, $\Omega_{m,0} = 0.3089$, and $\Omega_{b,0} = 0.0486$, which we adopt for this analysis.
TNG follows nine chemical elements: H, He, C, N, O, Ne, Mg, Si and Fe. Every stellar particle in the simulation inherits the gas elemental abundance out of which it formed and takes into account chemical enrichment via three distinct stellar evolutionary channels: AGB stars, core-collapse supernovae, and type Ia supernovae. The produced elements are released to the surrounding gas at the end of the stellar lifetime \citep[the detailed algorithm and yield tables used can be found in][section 2.3.3 and 2.3.4]{2018Pillepich}. Most importantly for this work, the dominant enrichment channel of magnesium and oxygen are core-collapse supernovae of massive stars ($8-100$~M$_\odot$), leading to an almost instantaneous enrichment due to their short lifetime. Iron enrichment happens to a large degree via type Ia supernovae, following a more extended delay time distribution that leads to a more gradual enrichment over time. In the gas phase, each cell contains information about the same nine elements, which are treated as passive scalars and advected with the gas flow without impacting it (other than indirectly via the radiative cooling rate, which is metallicity-dependent). This implies that the gas elemental abundance at the position of a star particle is simply given by the respective abundance of the gas cell in which this star particle is located. To convert mass fractions provided by these elemental abundances to relative metallicities, we adopt solar values from \citet{2009Asplund} and atomic weights from \citet{2021IUPAC}.

In this work we are analyzing the TNG100-1 simulation, i.e., the highest resolution run of the TNG100 volume.
TNG100 refers to the mid-sized simulation box with a volume of $110.7^3 \text{ Mpc}^3$. With such dimensions, the TNG100 box provides a balanced look into both large scale phenomena such as galaxy clustering and small scale properties like galaxy structure and gas composition. Using the highest resolution run with $2\times1820^3$ resolution elements, each galaxy considered in this study contains at least $2\times10^4$ star particles, sufficient to analyze not only their global properties but also their radial structure.

\subsection{Sample Selection} \label{subsec:sample selection}

\begin{figure}
    \centering
    \includegraphics[width=0.47\textwidth]{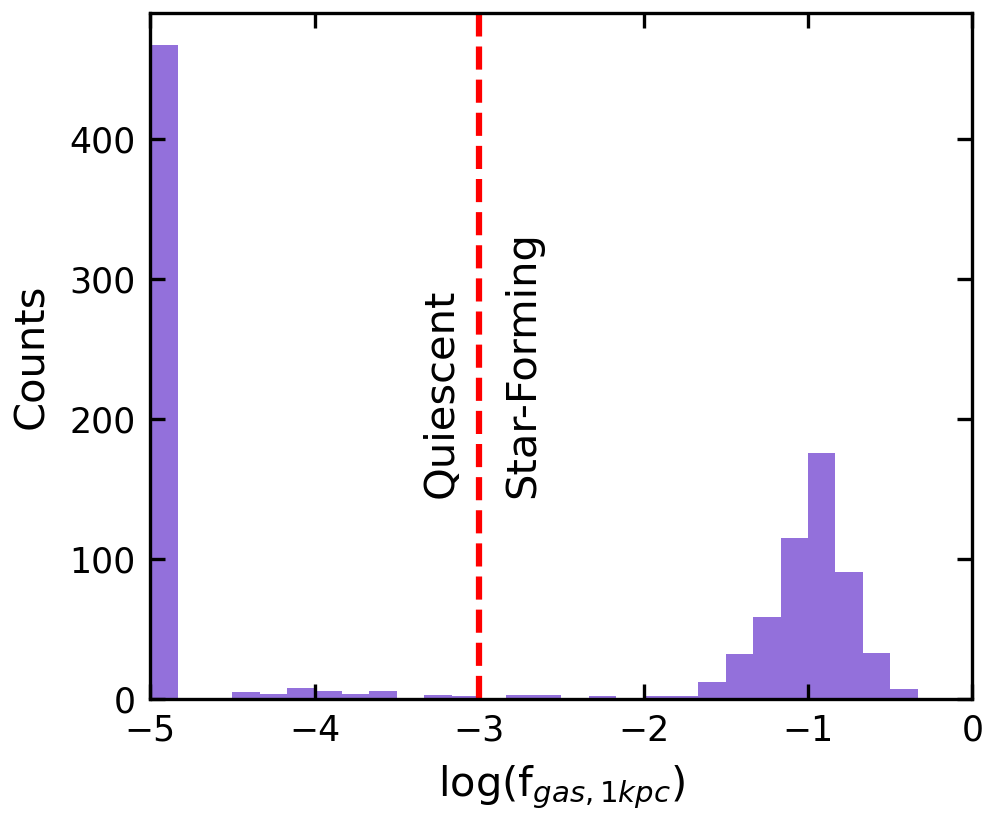}
    \caption{The bimodal distribution of central gas fraction for high-mass $z=2$ TNG galaxies. Outliers lower than $-5$ dex are automatically placed in the leftmost visible bin. These tend to have low gas density centers where a single cell of $\sim 10^6$~M$_\odot$ fills the entire inner kpc. The red dotted line at -3 dex marks the cutoff between star-forming (right) and quiescent (left) galaxies.}
    \label{fig:bimodal}
\end{figure}

We select $z=2$ galaxies with stellar masses $M_\star \geq 10^{10.5}M_\odot$, and obtain 1045 objects from the publicly available simulation data \citep{2019Nelson}.
For each galaxy we then calculate the gas fraction $f_\text{gas}$:
\begin{equation}
    f_\text{gas} \equiv \frac{M_\text{gas}}{M_\text{gas} + M_\star} \; ,
    \label{eqn:fgas}
\end{equation}
where $M_\mathrm{gas}$ is the gas mass and $M_\star$ is the stellar mass. We consider in particular $f_\text{gas, 1kpc}$, i.e., the gas fraction within the central 1~kpc. Figure~\ref{fig:bimodal} shows the distribution of $f_\text{gas, 1kpc}$ for the TNG galaxy population at $z=2$. The central gas fraction follows a clearly bimodal distribution, which we use to classify galaxies. We define quiescent galaxies as those with $\log f_\text{gas, 1kpc} \leq -3$ (505 systems at $z=2$), and as star-forming galaxies those with $\log f_\text{gas, 1kpc} > -3$ (540 systems). 

Our selection of quiescent systems is based on the central gas fraction instead of the more commonly adopted star formation rate because this choice leads to a greater similarity in radial gas profiles for star-forming galaxies, which will be discussed in Section~\ref{sec:profiles}. The galaxy-wide measurements, discussed in Section~\ref{sec:global_measurements}, do not qualitatively change when adopting a selection based on star formation rate. This is expected, since in TNG there is a close relation between central gas mass and star formation rate \citep[][]{TNG4, 2020Terrazas}.

In the TNG simulation, the central gas fraction is bimodal over the redshift range $0 < z < 4$, and we use the same selection to extract two comparison samples at lower redshifts. Using identical criteria in stellar mass and gas fraction, we select 2452 massive galaxies (of which 1618 are quiescent) at intermediate redshift, $z=0.7$, and 2855 massive galaxies (of which 2423 are quiescent) at $z=0$.

\section{Global Stellar Abundances}  \label{sec:global_measurements}

For the majority of galaxies at intermediate and high redshift, observational studies can only obtain global (i.e., galaxy-wide) measurements of stellar abundances. In this section we thus focus on the global stellar abundances of simulated galaxies, which we calculate by taking the luminosity-weighted average of all stellar particles within the effective radius (i.e., the radius containing half of the total stellar mass of the galaxy). We only consider the abundances of magnesium and iron, which are the elements that can be more easily measured in observations.

\subsection{Scaling Relations} \label{subsec:scaling relations}

\begin{figure*}
    \centering
    \includegraphics[width=0.8\textwidth]{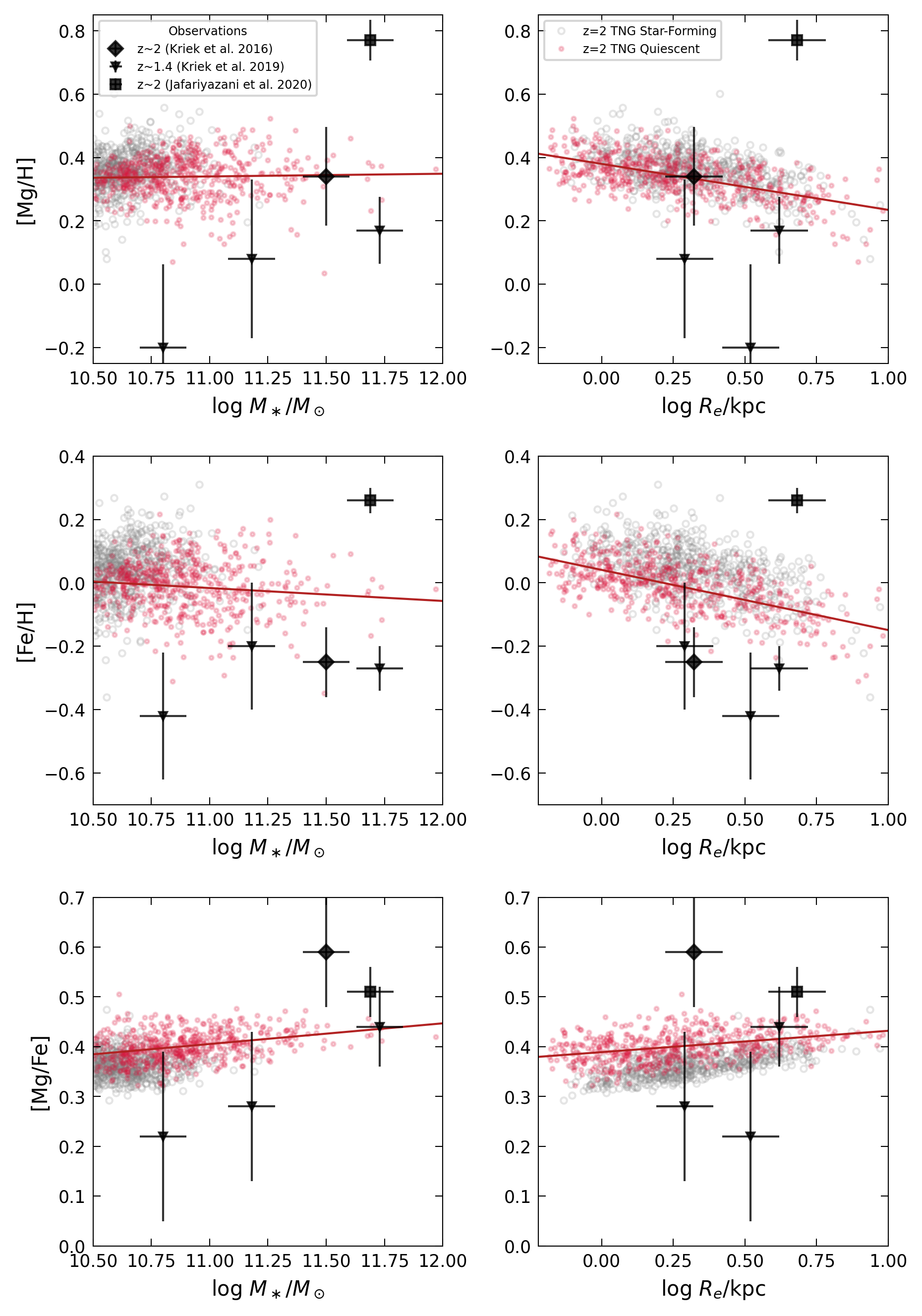}
    \caption{Stellar abundance ratios for high-redshift galaxies as a function of mass (left) and effective radius (right). TNG galaxies at $z = 2$ are shown as small circles, with quiescent systems in red and star-forming galaxies in gray. In each panel the red line is a linear fit to the simulated quiescent galaxies. Larger, black symbols mark observed galaxies \citep{2016Kriek, 2019Kriek, 2020Jafariyazani}. We adopt a stellar mass uncertainty of 0.1~dex for the observations, which is a lower limit on the typical statistical error \citep{2015Mobasher}.}
    \label{fig:z2_bigcheddar}
\end{figure*}

Figure~\ref{fig:z2_bigcheddar} shows [Mg/H], [Fe/H], and [Mg/Fe] as a function of galaxy mass (left) and effective radius (right). Each small circle represents an individual TNG galaxy at $z=2$; quiescent galaxies are shown in red, star-forming galaxies in gray. Red lines represent linear fits to the quiescent TNG galaxies. The TNG simulation makes clear predictions regarding the stellar abundances of massive galaxies at $z=2$:
\begin{itemize}

    \item The abundances of magnesium and iron are approximately flat as a function of stellar mass. This is analogous to the observed mass-metallicity relation in the local universe, which features a positive trend for low-mass systems but is mostly flat for $M_\star > 10^{10.5} M_\odot$ \citep{2005Gallazzi}.
    
    \item The relative abundance [Mg/Fe] increases slightly with stellar mass; however, all quiescent galaxies are found within a remarkably narrow range of values, with a spread of $\sim0.1$~dex.
    
    \item Both [Mg/H] and [Fe/H] feature a negative trend with the galactic effective size, with logarithmic slopes of -0.15 and -0.19, respectively. On the other hand, their ratio [Mg/Fe] follows a mildly positive trend with size (logarithmic slope: 0.04). These trends will be the subject of further investigation in Section~\ref{sec:profiles}.
    
    \item Star-forming galaxies tend to be slightly more enriched than quiescent galaxies both at fixed mass and at fixed radius, especially in [Fe/H]; they also have lower [Mg/Fe]. This is consistent with their star formation histories being more extended, thus providing more time for type Ia supernovae to go off and deposit iron in the interstellar medium prior to the next generation of stars being formed.

\end{itemize}

In Figure~\ref{fig:z2_bigcheddar} we also include high-redshift observations from \citet{2016Kriek}, \citet{2019Kriek}, and \citet{2020Jafariyazani}, shown as large black markers with error bars. All these observational studies use similar types of data and identical spectral fitting techniques to derive stellar abundances. 
The simulated galaxy population is in broad agreement with the observational data, but the sparse sampling and large uncertainties of the observations prevent a detailed comparison. One notable exception is the ultra-massive galaxy observed by \citet{2020Jafariyazani}, whose stellar abundance is significantly larger than what we found in the simulation, particularly for [Mg/H]. Moreover, using a different method based on spectral indices, \citet{2020Lonoce} measured a wide range of abundances (spanning from -0.8 to +0.6) for a sample of four massive galaxies at high redshift. We note, however, that at these high masses the TNG galaxy sample is incomplete due to limitations in the simulated volume.

\begin{figure*}
    \centering
    \includegraphics[width=\textwidth]{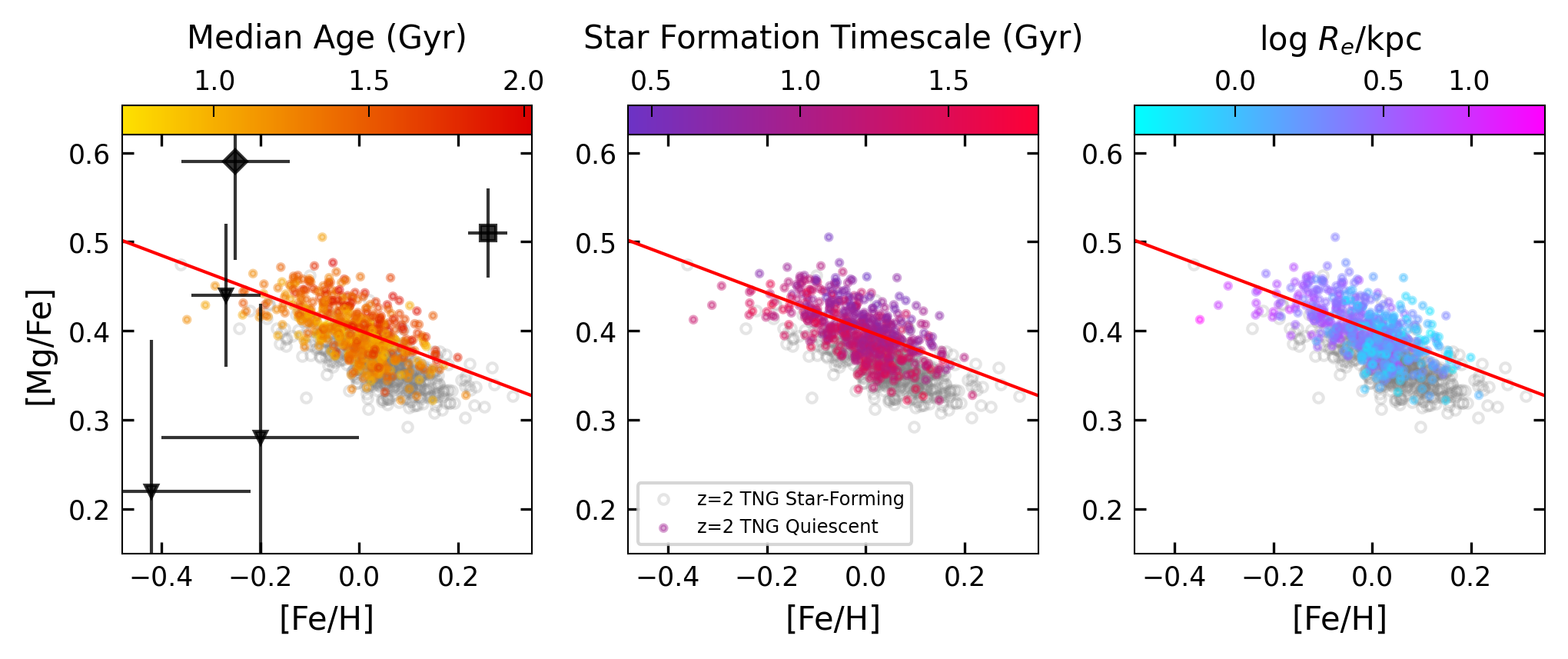}
    \caption{[Mg/Fe] vs [Fe/H] for $z = 2$ simulated galaxies. In each panel, quiescent galaxies are color-coded according to a different variable: the median stellar age, the star formation timescale (both calculated within one effective size), and the effective size. In the left panel we also show the observations at $z\sim2$ \citep{2016Kriek, 2019Kriek, 2020Jafariyazani}, following the legend in Figure~\ref{fig:z2_bigcheddar}. The red line is a linear fit to the quiescent simulated galaxies.}
    \label{fig:z2_three}
\end{figure*}

\begin{figure*}
    \centering
    \includegraphics{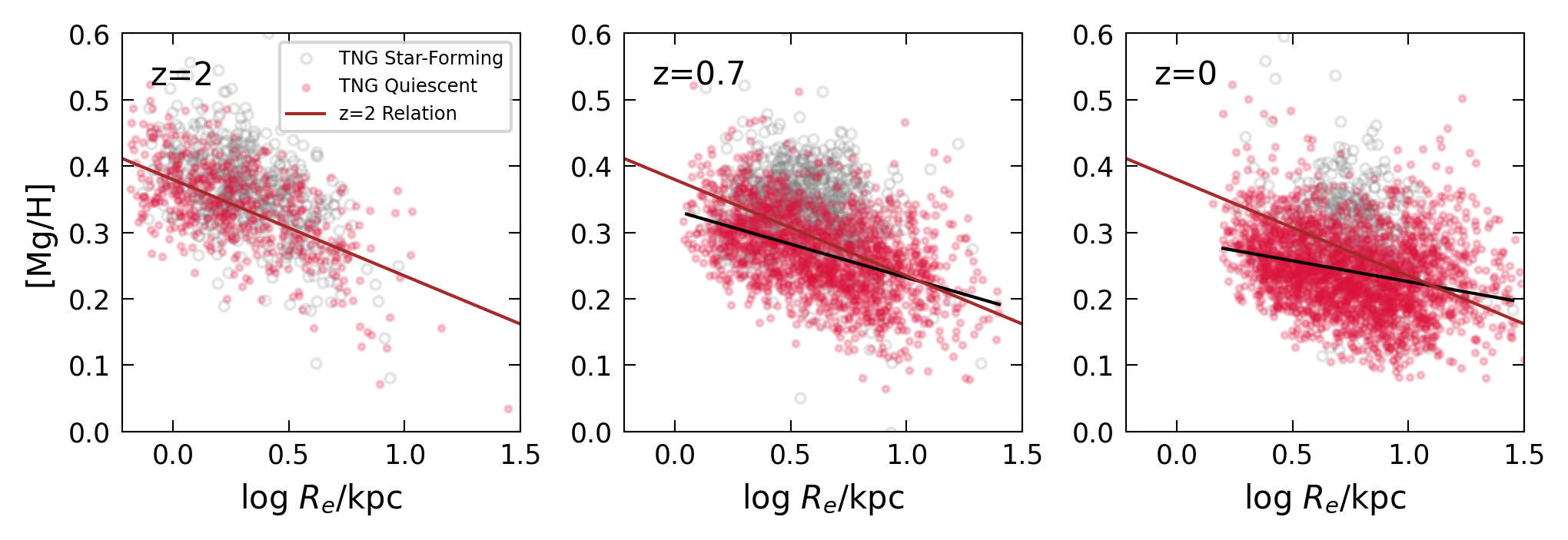}
    \caption{Stellar [Mg/H] vs effective size for simulated galaxies, in three redshift bins. The red line, seen in all panels, is a linear fit to the $z=2$ quiescent systems. The black lines in the last two panels are linear fits to the quiescent galaxies at their respective redshift.}
    \label{fig:zall_MgHRe}
\end{figure*}

Given the difficulty of comparing the TNG predictions to the observations at $z\sim2$, we also perform the comparison at lower redshift. In Appendix~\ref{app:0.7scalings}, we show the TNG abundance relations for the $z=0.7$ snapshot (which are qualitatively similar to those found at $z=2$), compared to the abundances measured by \citet{2021Beverage} for galaxies observed at the same redshift. The observed [Mg/H] and [Fe/H] scaling relations are systematically offset compared to the simulations, but feature similar slopes. In particular, the observed abundances of quiescent galaxies appear to follow a flat trend with mass and a negative trend with radius. The anti-correlation between stellar abundances and galaxy sizes has also been pointed out by several observational studies carried out at low and intermediate redshift \citep{2015McDermid, 2018Barone, 2021Beverage, 2022Barone}. The relative abundance [Mg/Fe], on the other hand, does not show appreciable trends in the observations, and features only a mild trend with mass in the simulations.

\begin{figure*}
    \centering
    \includegraphics{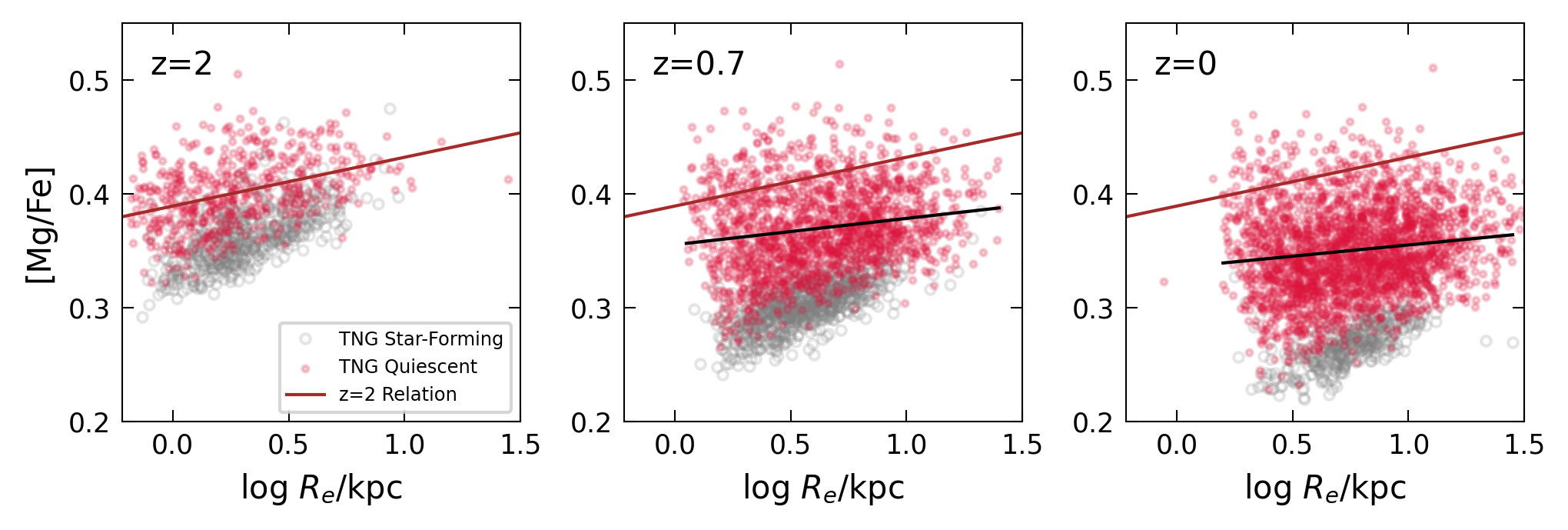}
    \caption{Stellar [Mg/Fe] vs effective size for simulated galaxies, in three redshift bins. The red line, seen in all panels, is a linear fit to the $z=2$ quiescent systems. The black lines in the last two panels are linear fits to the quiescent galaxies at their respective redshift.}
    \label{fig:zall_MgFeRe}
\end{figure*}

\begin{figure*}
    \centering
    \includegraphics{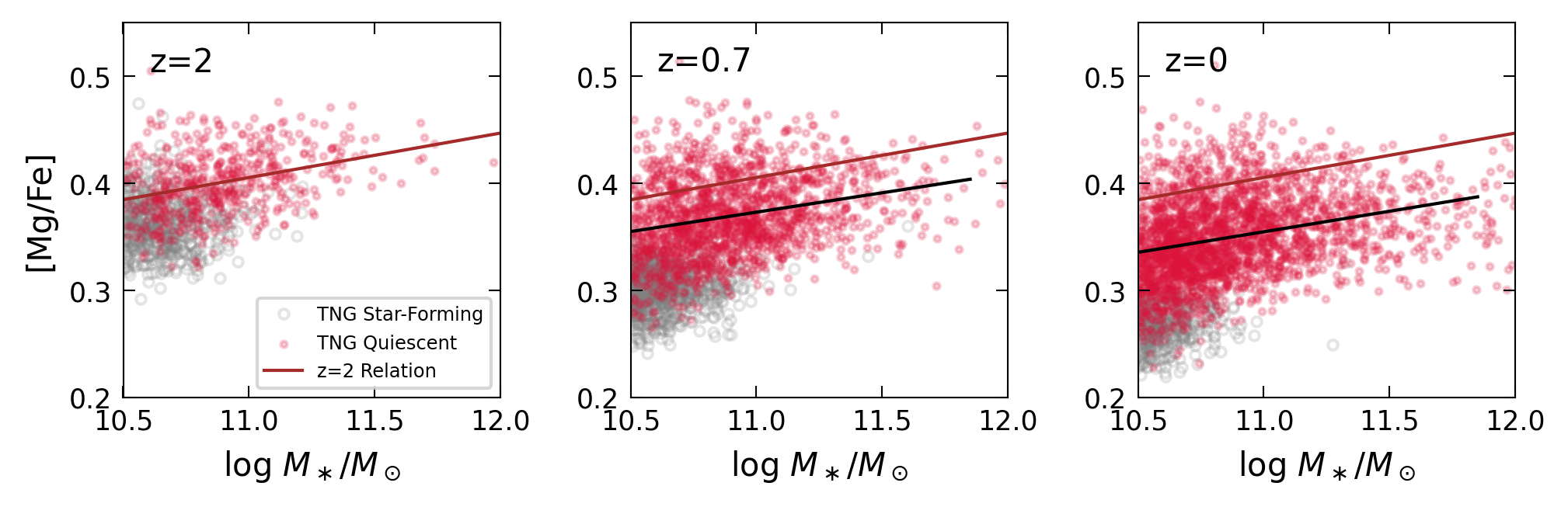}
    \caption{Stellar [Mg/Fe] vs stellar mass for simulated galaxies, in three redshift bins. The red line, seen in all panels, is a linear fit to the $z=2$ quiescent systems. The black lines in the last two panels are linear fits to the quiescent galaxies at their respective redshift.}
    \label{fig:zall_MgFeMass}
\end{figure*}

\subsection{Abundance Patterns} \label{subsec:abundance_patterns}

The relation between [Mg/Fe] and [Fe/H] is often used to probe the timescale of galaxy formation with the aid of chemical models \citep[e.g.,][]{2016Kriek}. In the simplest scenario, when galaxies form rapidly their abundance pattern is mostly set by the material produced by short-lived massive stars, and thus features high [Mg/Fe] and low [Fe/H]. In contrast, in the case when systems form slowly, there is sufficient time for low-mass stars to affect the composition of later stellar generations. As a result, [Mg/H] is low and [Fe/H] is high. We show the abundance patterns of simulated galaxies in Figure~\ref{fig:z2_three}; in each panel we color code the quiescent systems according to the median stellar age, the formation timescale (defined as the difference between the 90th and 10th percentile of the stellar age distribution), and the effective size, respectively. The first panel also includes the same observations as in Figure~\ref{fig:z2_bigcheddar}.
Across all panels, the same fit to the [Mg/Fe] abundance as a function of [Fe/H] is shown in red, highlighting a relatively tight negative correlation. This trend is apparently consistent, at least qualitatively, with the prediction of the simple chemical models described above. However, the color coding reveals that this anti-correlation between [Mg/Fe] and [Fe/H] is \emph{not} driven by changes in formation timescale or median age. 
The effect of the formation timescale is mostly vertical, with galaxies that form rapidly being more enriched in [Mg/Fe] at fixed [Fe/H]. Stellar ages have an analogous effect but in the opposite direction, consistent with the fact that galaxies that form slowly are, on average, younger. The star-forming population is found at the lower [Mg/Fe] envelope of the overall distribution, again consistent with having the youngest stellar ages and longest star formation timescales.
The strongest trend along the sequence is, instead, with the effective size: smaller galaxies have systematically lower [Mg/Fe] and higher [Fe/H]. Thus, the sequence that we see in this diagram is tightly linked to the trends with effective size previously shown in Figure~\ref{fig:z2_bigcheddar}.

In Appendix~\ref{app:0.7scalings} we analyze the abundance patterns for simulated galaxies at $z=0.7$, and compare them to the available observations.

\subsection{Redshift Evolution}
\label{sec:redshift_evolution}

Next, we explore the galaxy-wide stellar abundances as a function of redshift. 
While iron is often taken as the reference element for historical reasons, from now on we will focus on the magnesium abundance, which is set by core-collapse supernovae and thus can be easily interpreted in the context of the instantaneous recycling approximation \citep[see, e.g., the discussion in][]{weinberg19}. Moreover, magnesium traces the total metallicity [Z/H] much more closely than iron, as can be seen by taking the \citet{thomas03} relation: [Z/H] = [Fe/H] + 0.94~[Mg/Fe], and rewriting it as [Z/H] = 0.06~[Fe/H] + 0.94~[Mg/H]. Accordingly, we will use ``stellar metallicity'' to indicate the magnesium abundance.

Figure~\ref{fig:zall_MgHRe} shows [Mg/H] as a function of effective size for simulated galaxies at $z=2$ (left), $z=0.7$ (center) and $z=0$ (right). All plots include the $z=2$ quiescent trend line for reference. Black lines on the right two panels mark the trends of quiescent galaxies in their respective redshifts.
The anti-correlation between [Mg/H] and effective size remains in place from $z=2$ to $z=0$, but with a shallower slope and a larger scatter at later cosmic times. Toward low redshift, massive galaxies also grow larger in size, as indicated by the rightward migration of the sample. The star-forming galaxy population tends to have slightly higher [Mg/H] values than the quiescent population for all three redshifts.

\begin{figure*}
    \centering
    \includegraphics[width=0.8\textwidth]{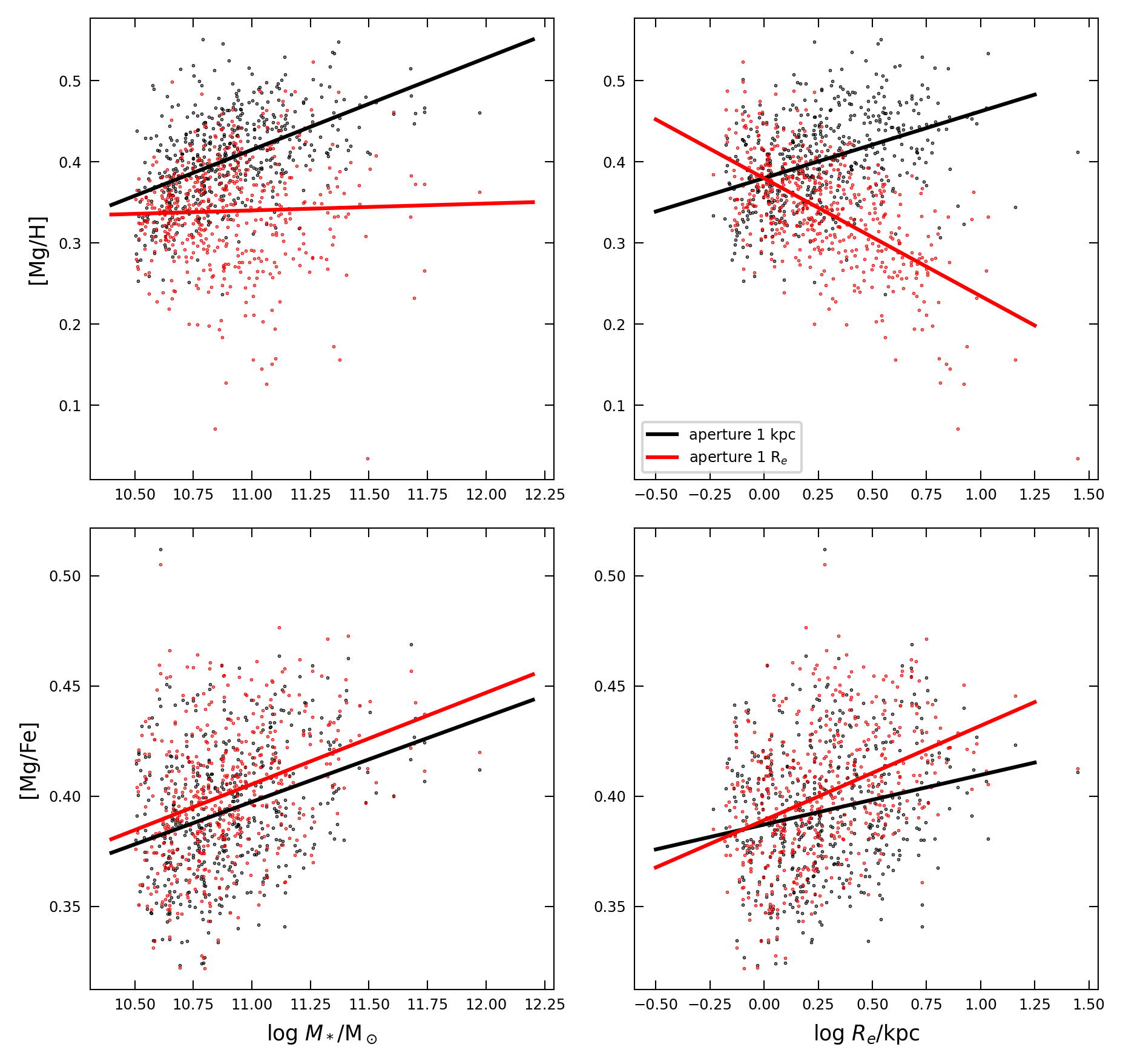}
    \caption{Abundance ratio measured in different apertures vs. stellar mass (left) and vs. effective radius (right) for quiescent $z=2$ galaxies. Red points show the abundances calculated within 1 effective size (like in the previous figures), while black points show the abundances calculated within a fixed aperture of 1 kpc. Solid lines represent linear fits to the data.}
    \label{fig:aperture_dependence_reff}
\end{figure*}

The relation between [Mg/Fe] and effective radius is shown in Figure~\ref{fig:zall_MgFeRe} for different redshifts. This relation becomes more shallow and its scatter increases with cosmic time, similar to what is seen for [Mg/H]. Overall, the clear trend with effective radius seen at $z=2$ is significantly weakened by $z=0$. Star-forming galaxies tend to have lower [Mg/Fe] values than quiescent galaxies at the same redshift, implying that they are more enhanced in [Fe/H] than in [Mg/H] relative to the quiescent population. While the average value of [Mg/Fe] in the star-forming population decreases towards lower redshift, they always make up the lower envelope of the overall galaxy population. 
Finally, Figure~\ref{fig:zall_MgFeMass} shows that the decrease in average [Mg/Fe] with cosmic time is a mass-independent effect, since the slope of the [Mg/Fe] vs. stellar mass relation does not appear to evolve with redshift. Figure~\ref{fig:zall_MgFeMass} further shows that the lower [Mg/Fe] in the star-forming population is also present at fixed stellar mass.

In conclusion, the scaling relations of simulated galaxies are substantially tighter at high redshift than at $z=0$. This likely reflects the importance of processes such as mergers, which can mix the stellar populations of different galaxies. Even in the absence of external processes, the abundance evolution of the quiescent galaxy population is affected by quenching: at each redshift, newly formed quiescent galaxies are added to the population, and their abundance follows the abundance distribution of star-forming galaxies at that redshift \citep[see, e.g.,][]{2020Choi}. As a result of this process, the abundance distribution widens towards lower redshifts. The $z=2$ relations are thus more ``pristine,'' and offer a cleaner way to both probe the physics of galaxy formation and test the TNG model using observations.

Finally, we caution that Figures~\ref{fig:zall_MgHRe}--\ref{fig:zall_MgFeMass} show different samples at different redshifts. As a result, the evolution of the scaling relations is due to a combination of physical evolution and a change in the sample composition, making it difficult to develop a simple, meaningful interpretation. For example, Figure~\ref{fig:zall_MgHRe} shows that the typical [Mg/H] abundance decreases with cosmic time, at fixed radius. However, this is mostly due to the contribution of metal-poor galaxies that enter the mass selection at low redshift: when we follow the $z=2$ sample forward in time, we find that the [Mg/H] vs. effective size relation \emph{increases} with time. Moreover, as each individual galaxy becomes bigger due to satellite accretion, the abundance measurement probes a different physical region of the galaxy, further complicating the interpretation. This effect is particularly important when interpreting the star-forming population, which can be considered as being in a transition state (entering the selection via stellar mass growth and leaving it via quenching, which is a one-way process for most galaxies).

\subsection{Aperture Dependence} \label{subsec:aperture_dependence}

One crucial aspect of stellar abundances, that must be carefully considered when comparing simulations with observations, is the aperture over which the abundances are measured. We test for this effect in Figure~\ref{fig:aperture_dependence_reff}, where we show the abundances of simulated galaxies calculated using two different apertures. The red symbols show the luminosity-weighted average abundance of all stars within the central effective radius (which is the aperture used to generate all previous figures), while the black symbols show the luminosity-weighted average abundance of all stars within the central kpc. For simplicity we restrict ourselves to the quiescent population at $z=2$.

The choice of the aperture is clearly important for the resulting scaling relations: The left panels of Figure~\ref{fig:aperture_dependence_reff} show that the abundance in the central kpc features a significant increase with stellar mass, as opposed to the flat relation that we find when using the abundances calculated within the central effective size.
Most strikingly, the right panels of Figure~\ref{fig:aperture_dependence_reff} show that the abundance trend with radius is flipped, switching from a negative to a positive slope when changing the aperture from 1 $R_e$ to 1 kpc.
We note that the trend for a measurement within a fixed aperture of $30$~kpc (not shown here) yields similar slopes than when using the effective radius.
The relative abundance [Mg/Fe], on the other hand, does not depend strongly on the aperture used.

\section{Radial Profiles}
\label{sec:profiles}

In this section, we move beyond galaxy-wide measurements and explore the spatial distribution of metal abundances within simulated galaxies. This is necessary in order to understand the aperture dependence of the scaling relations, and to investigate the physical origin of the abundance relations, including the connection between stellar and gas-phase metallicity. 

\subsection{Stellar Abundance Profiles}
\begin{figure}
    \centering
    \includegraphics{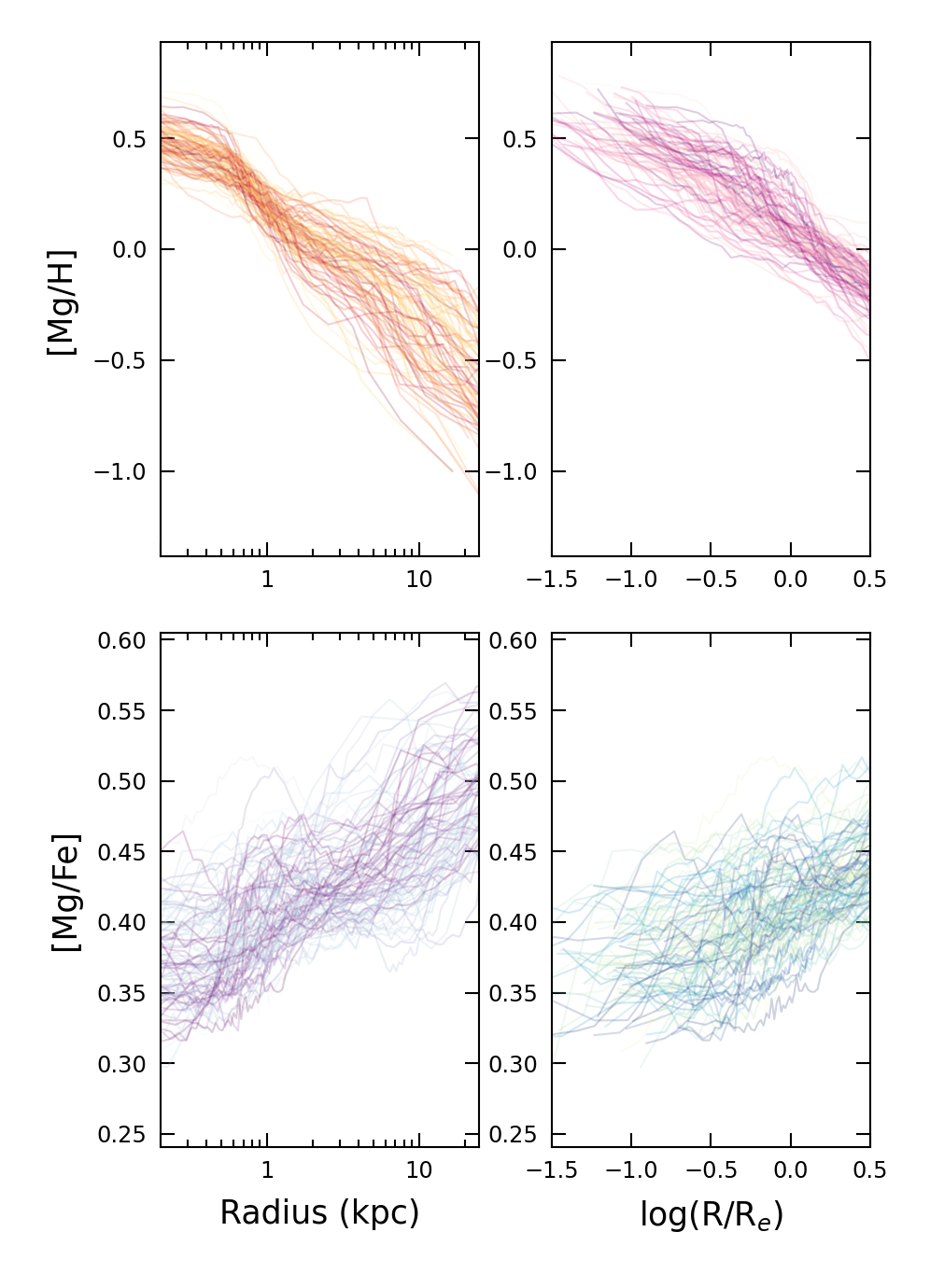}
    \caption{Radial profiles of stellar [Mg/H] (top) and [Mg/Fe] (bottom) for 100 random quiescent TNG galaxies at $z=2$. The profiles are shown both as a function of physical radius in kpc (left) and as a function of radius normalized by the galaxy effective size (right). Note that the bottom panels span a much smaller dynamic range.}
    \label{fig:profiles}
\end{figure}

Figure~\ref{fig:profiles} shows the stellar abundance profiles of 100 randomly chosen simulated quiescent galaxies at $z=2$ as a function of radius, expressed in units of kpc (left) and in units of the effective radius (right). The top panels show [Mg/H] and the bottom panels show [Mg/Fe].  
The [Mg/H] profiles are steeply declining, spanning a dynamic range of roughly 1~dex. The [Mg/Fe] profiles, in contrast, are rising and shallow, spanning only $\sim0.1$~dex. The scatter in the bottom panels appears to be larger because of the different scale, while in reality it is much smaller. 
Interestingly, the scatter in the [Mg/H] profiles in the inner parts is smaller when plotted against radius in kpc instead of in units of the effective size. This indicates that the process setting the abundance profile is different from the process responsible for setting galaxy sizes. Note, however, that in the outskirts (outside of $\sim3$~kpc or $\sim1$ effective radius) the opposite seems to be the case.

The steepness of the [Mg/H] profiles, together with the comparably low scatter of the profiles as a function of physical radius, explain the aperture dependence of the scaling relations discussed in Section~\ref{subsec:aperture_dependence}. When considering the abundance within the effective size, small galaxies only ``probe'' the central, enriched part of this quasi-universal profile, while larger galaxies ``probe'' a substantial number of distant and metal-poor stars, bringing down the luminosity-weighted abundance average by $0.1$ to $0.2$~dex. On the other hand, measuring the enrichment in an aperture of $1$~kpc only probes the galactic center, which differs from the galaxy-wide measurement in the case of more extended galaxies, thus altering the slope of the [Mg/H] vs. effective radius relation. The [Mg/Fe] ratio suffers less from this effect since the gradient is less steep. 

What remains to be explored is the origin of the steepness, the small scatter, and the seemingly universal stellar metallicity profile in the inner regions of $z=2$ galaxies, for which we have to turn toward the connection between gas enrichment and stellar metallicity. 

\subsection{Gas vs. Stellar Metallicity}

\begin{figure*}
    \centering
    \includegraphics{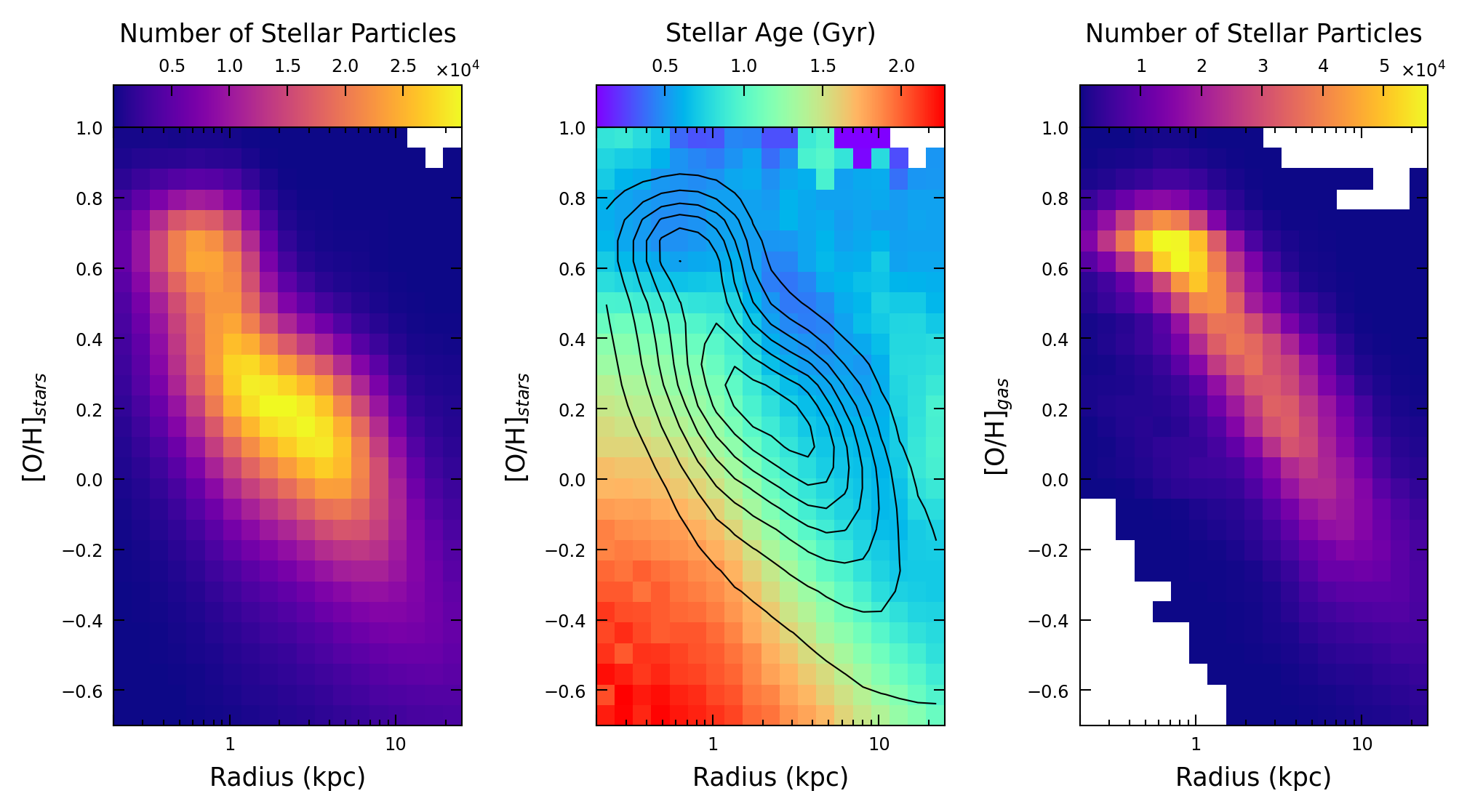}
    \caption{Metallicity vs radius for 100 random star-forming TNG galaxies at $z=2$. \emph{Left:} The color coding shows the number of stellar particles (belonging to any galaxy) in each bin of stellar [O/H] and galactocentric radius. \emph{Center: } The color coding shows the median stellar age in each bin; the number of stellar particles shown in the previous panel is overlaid in black. \emph{Right:} Same as the first panel, but for gas-phase [O/H].}
    \label{fig:OHage_starforming}
\end{figure*}

Since stars inherit the metallicity of the gas from which they form, we need to explore the gas-phase abundance if we want to understand the origin of the stellar abundance profiles. However, quiescent galaxies are devoid of gas (at least in the center) by selection; we thus turn to the massive, $z=2$ star-forming galaxies as examples of progenitor galaxies, and study their gas properties. From now on, we focus on the abundance profiles as a function of physical radius in units of kpc, in light of the lower scatter.
Moreover, to remain as close as possible to the observations, we will consider the abundance of oxygen instead of magnesium. While they are both $\alpha$ elements, the oxygen abundance is much easier to measure for the gas, particularly at high redshift, and is often taken as the ``gas metallicity.'' We checked that the stellar [Mg/H] and [O/H] profiles are virtually identical so that the conclusions we draw from analyzing the oxygen abundance can also be applied to magnesium.

We compare gas and stellar abundances for 100 random star-forming galaxies at $z=2$ in Figure~\ref{fig:OHage_starforming}. 
The first panel shows the stacked [O/H] abundance profile for all 100 star-forming galaxies: the color coding represents the number of stellar particles (belonging to any galaxy) that fall within each two-dimensional bin. 
The middle panel is colored according to the median age of star particles in each bin, ranging from 0 to 2.5 Gyr. Black contour lines overplot the particle density as seen in the first panel. The overall abundance profile appears to be due to the superposition of stellar populations of different ages, each featuring an abundance profile with roughly the same slope but varying normalization from metal-poor for the oldest stars to metal-rich for the youngest ones. The third panel shows the gas-phase metallicity profile, which is very similar to the distribution of the youngest stars seen in the second panel.
This confirms that the slope of the stellar abundance profile is primarily set by the slope of the gas-phase abundance profile. However, the stellar abundance profile is also modified by different formation times of stars at different radii and its scatter at fixed radius, which, due to a time evolving gas metallicity profile, causes a larger scatter in stellar metallicities at fixed radius as well as additional features in the profile.
Therefore, the steepness of the metallicity gradients in quiescent galaxies is a remnant of steep gas-phase metallicity gradients during their star-forming phase (assuming little radial mixing has occured, which is a reasonable assumption at $z=2$). The low galaxy-to-galaxy variation of stellar metallicity further indicates that the gas metallicity profiles during the star-forming phase were very similar as well.

\subsection{Gas Fraction and Effective Yield}

\begin{figure*}
    \centering
    \includegraphics{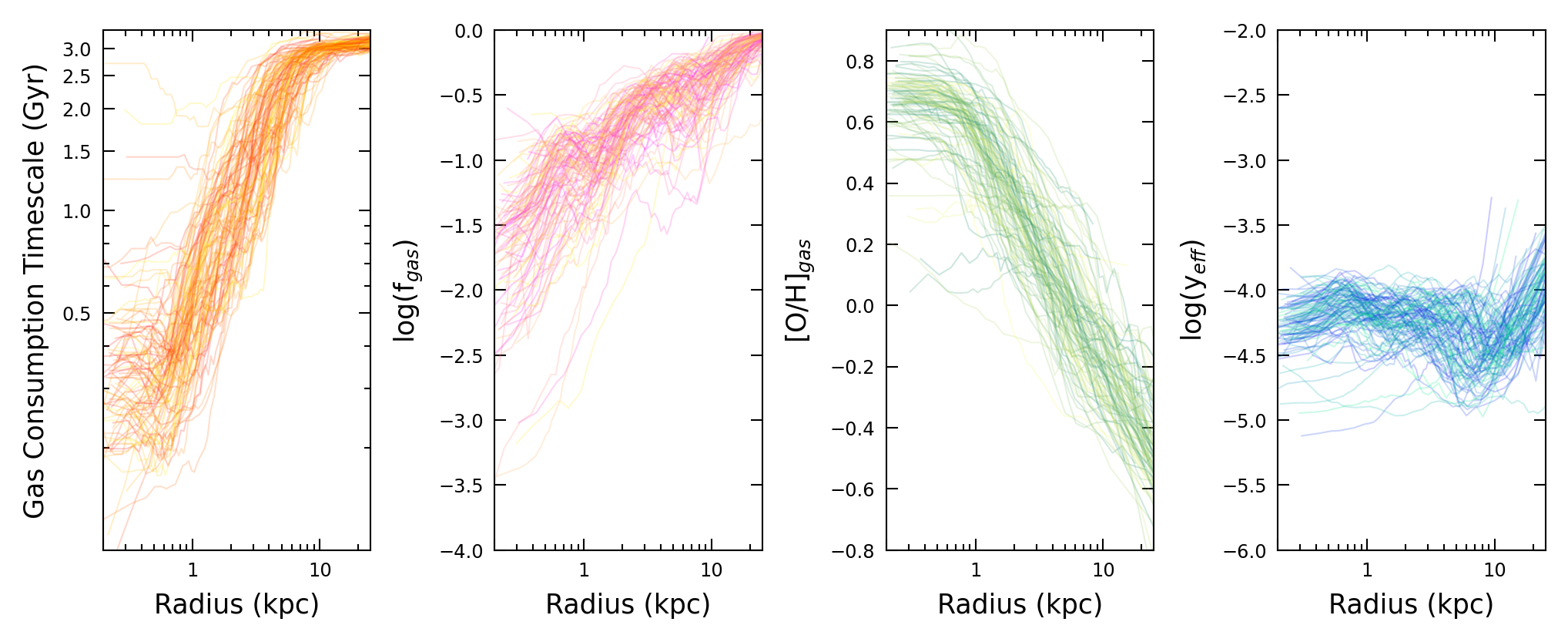}
    \caption{Radial profiles of gas properties for the same random 100 star-forming TNG galaxies at $z=2$ shown in Figure~\ref{fig:OHage_starforming}. From left to right, the panels show the radial profiles for the gas consumption timescale (i.e., local gas mass divided by local star formation rate), gas fraction, gas-phase oxygen abundance, and effective yield as defined in Equation~\ref{eqn:yeff}.}
    \label{fig:yeff}
\end{figure*}

To further understand the origin of the gas metallicity profile, in Figure~\ref{fig:yeff} we explore the properties of the gas in the sample of simulated star-forming galaxies.
The first panel shows the radial profiles of the gas consumption timescale, defined as the ratio between the local gas mass density and the local star formation rate density. There is a clear spatial variation, with the galaxy centers being efficient at forming stars (with timescales of a few hundred Myr), and the galaxy outskirts being inefficient, taking about 3 Gyr to transform the gas into stars --- a time comparable to the age of the universe at $z=2$. As a consequence, the gas fraction (shown in the second panel) is only a few percent in the center, where most of the gas has already been transformed into stars (or removed by galactic outflows), but approaches 100\% in the outskirts.

The third panel of Figure~\ref{fig:yeff} shows the gas-phase abundance profile, which is clearly anti-correlated with both the consumption timescale and the gas fraction: regions where most of the gas has been transformed into stars feature a high metallicity, while regions where gas is transformed very slowly are metal-poor. This behavior is in qualitative agreement with the expectation of the simplest model of chemical evolution: the closed-box system. In this model, an initial mass of metal-poor gas is transformed into stars without further inflows or outflows. The model is characterized by a simple relation between the gas metallicity and the gas fraction, which depends only on the effective yield, defined as
\begin{align}
    y_\text{eff} \equiv \frac{Z_\text{gas}}{\ln(1/f_\text{gas})} \; ,
    \label{eqn:yeff}
\end{align}
where $Z_\text{gas}$ is the fraction of the total baryonic mass contributed by a specific metal. In our case we consider oxygen, so $Z_\text{gas} = 10^\text{[O/H]} \cdot (\text{M}_\text{O} / \text{M}_\text{tot})_\odot$.
%, with $(\text{M}_\text{O} / \text{M}_\text{tot})_\odot = 0.0058$ \citep{2009Asplund}.
We show the radial profile of the effective yield in the last panel: for most galaxies, this quantity remains approximately constant at around $10^{-4}$, out to $\sim10$ kpc. 
%I believe that the effective yield is much lower than the actual oxygen yield input in TNG, meaning that a large fraction of metals must escape, thus it is most definitely not a closed box.
This is a remarkable result: while the gas fraction and the gas abundance vary by more than an order of magnitude along the galactocentric radius, the effective yield, which is a simple combination of these two quantities, is roughly constant. Such result suggests that galaxies are, indeed, similar to closed-box systems \emph{at fixed radius}.
The local closed-boxed systems at two different radii have the same effective yield, but due to their different star formation efficiencies they are in two different phases of evolution, characterized by different values of gas fraction and metallicity.

\begin{figure*}
    \centering
    \includegraphics{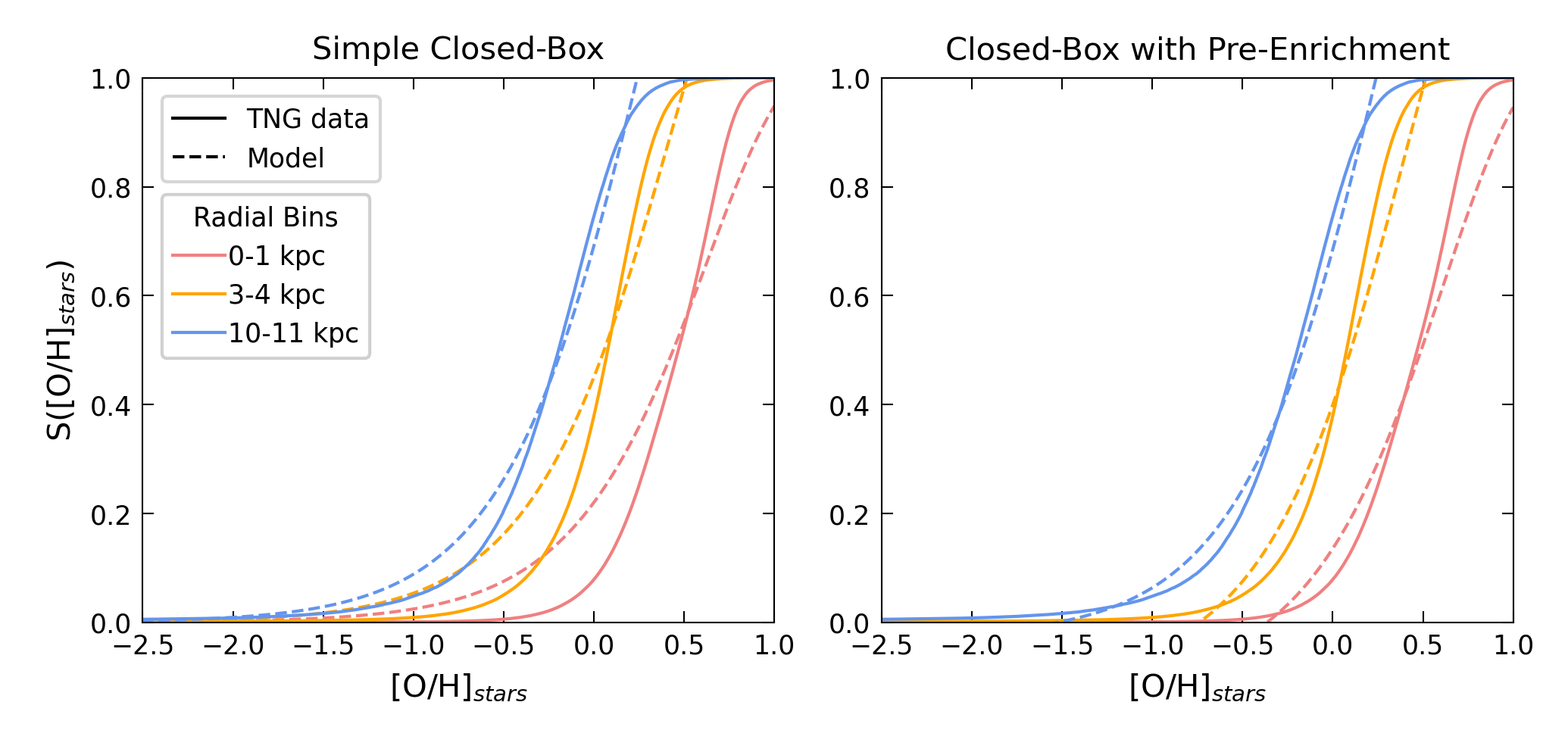}
    \caption{The cumulative distribution function, describing the fraction of stars below a given value of [O/H]. 
    In both panels, the solid lines show the distribution function measured for 100 star-forming TNG galaxies at $z=2$ in three different radial bins.
    The dashed lines show the best fit for two toy models: the simple closed box in the left panel, and the pre-enriched closed box in the right panel. Note the improved fit at low metallicities once pre-enrichment is included.}
    \label{fig:CDF_big}
\end{figure*}

\begin{figure}
    \centering
    \includegraphics{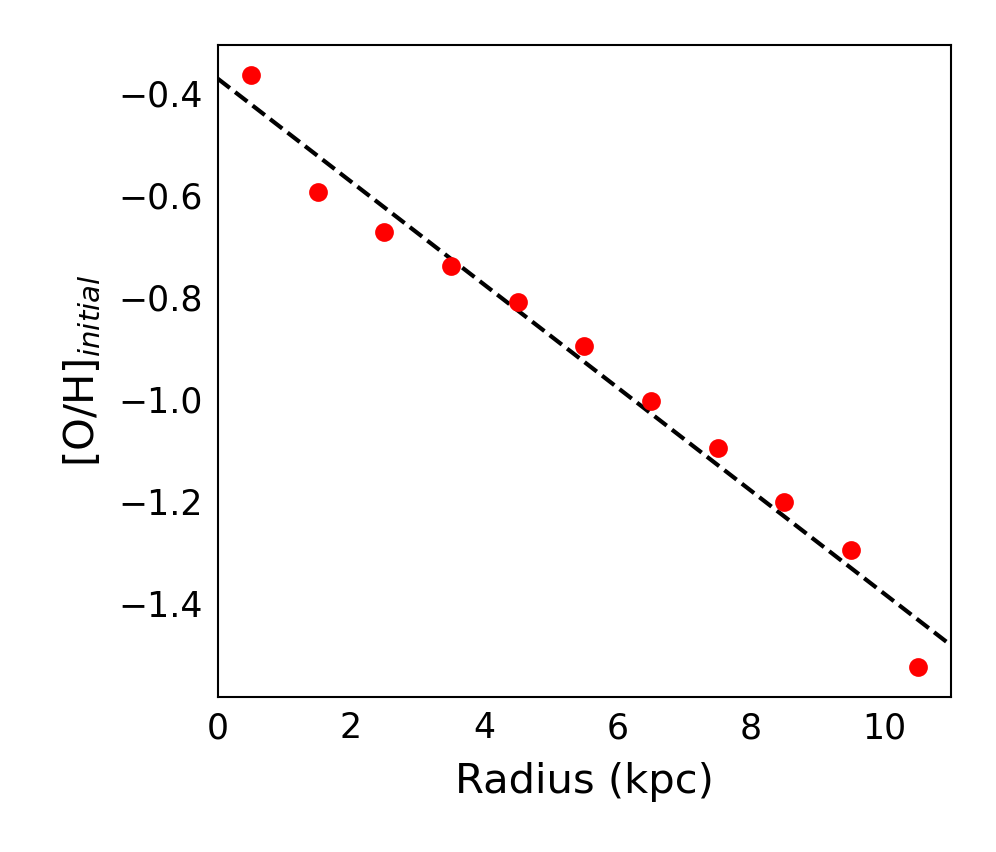}
    \caption{The best-fit pre-enrichment abundance for 100 star-forming TNG galaxies at $z=2$, based on Equation \ref{eqn:S(Z)_new}. Each red point represents the center of a 1-kpc radial bin. The linear fit, shown in black, has a slope of approximately -0.10 dex/kpc.}
    \label{fig:CDF_z0}
\end{figure}

\subsection{Closed-Box Model} \label{sec:closedbox}

Motivated by the approximate flatness of the effective yield profile shown in Figure~\ref{fig:yeff}, we construct a toy model in which each radial shell of a galaxy is a closed-box system. We also assume the instantaneous recycling approximation (that is, there is no delay between star formation and the return of enriched material to the interstellar medium), which is appropriate for $\alpha$ elements such as magnesium and oxygen. In this form, the closed-box system has been extensively studied, and analytical solutions exist \citep[see][for a review]{2012Matteucci}.

We test our toy model by exploring the distribution of stellar abundances. The closed-box model predicts a simple function for $S(Z)$, the fraction of stars with metallicity lower than $Z$:
\begin{equation}
    S(Z) = \frac{1 - (f_\text{gas})^{Z/Z_\text{final}}}{1 - f_\text{gas}} \; .
    \label{eqn:S(Z)}
\end{equation}
The metallicity distribution of a closed-box system thus depends only on the current gas fraction $f_\text{gas}$ and the final stellar metallicity $Z_\text{final}$, which coincides with the current gas-phase metallicity and represents the maximum possible value for the stellar metallicity. In the left panel of Figure~\ref{fig:CDF_big}, we compare this prediction to the stellar oxygen abundances measured for star-forming TNG galaxies in three radial bins, each with a width of 1 kpc. For each bin we adopt the median value of the measured $f_\text{gas}$, and then apply a least squares fit to find $Z_\text{final}$. Fit values are listed in Table~\ref{tbl:CDF}; note that in the table and figures we convert all mass fractions $Z$ to the corresponding logarithmic abundances [O/H]. The model is able to match the median [O/H] in each radial bin, and broadly reproduces the shape of the cumulative distribution of oxygen abundance. This means that the toy model can explain the overall trend of the stellar metallicity profiles. However, we note two important discrepancies. First, the fits return values of [O/H]$_\text{final}$ that are 0.3--0.4 dex larger than the gas-phase abundance measured in the same bins (see Table~\ref{tbl:CDF}). This suggests that the gas metallicity was higher at some point in the past, which is impossible in a closed-box system. Second, our simple model produces too many stars with low metallicity compared to the TNG simulation (see Figure~\ref{fig:CDF_big}, left panel). This is analogous to the discrepancy between the closed-box model prediction and observations of stellar metallicities in the Milky Way \citep[e.g.,][]{1980Tinsley}. One simple way to avoid this discrepancy is to assume that the initial gas is pre-enriched, so that even the earliest generations of stars are relatively metal-rich. In this case, the expected distribution of stellar metallicity becomes
\begin{equation}
    S(Z) = \frac{1 - (f_\text{gas})^{(Z-Z_\text{initial})/(Z_\text{final}-Z_\text{initial})}}{1 - f_\text{gas}} \; ,
    \label{eqn:S(Z)_new}
\end{equation}
where $Z_\text{initial}$ is the metallicity of the gas at the beginning of the closed-box evolution. Using this equation, and adopting the same $f_\text{gas}$ and $Z_\text{final}$ found above, we fit the TNG metallicities for $Z_\text{initial}$ in each radial bin and show the results in the right panel of Figure~\ref{fig:CDF_big}. Clearly, this model improves our prediction of the TNG stellar abundances. We repeat the fitting procedure in several 1 kpc radial bins, and plot the resulting profile of the initial gas abundance in Figure~\ref{fig:CDF_z0}. The pre-enrichment value is about two-fifths of the solar metallicity (-0.4 dex) in the galaxy center and declines with a roughly constant gradient of $-0.10$ dex per kpc.

In conclusion, we find that the \emph{gas-phase} metallicity as a function of radius can be modeled by a series of closed-box shells with identical effective yield. However, such a model is too simplistic to describe the distribution of \emph{stellar} metallicity, which can be explained with closed-box models only if one allows a radially varying pre-enrichment. The pre-enrichment model is also very simplistic, and represents only one of the several possible solutions. Given the discrepancies arising from the use of a closed box model to describe the TNG data, we conclude that gas inflows and outflows are likely to play an important role in determining the stellar abundance profiles in TNG.

\begin{table}[]
\centering
\begin{tabular}{ccccc}
\hline
Radial bin & $f_\text{gas}$ & {[O/H]$_\text{gas}$} & {[O/H]$_\text{final}$} & {[O/H]$_\text{initial}$} \\ 
(kpc) &   &   &   &   \\ 
\hline
{[}0, 1{]}       & 0.045     & 0.643      & 1.119              & -0.364 \\
{[}3, 4{]}       & 0.271     & 0.192      & 0.516              & -0.739 \\
{[}10. 11{]}     & 0.370     & -0.124      & 0.240             & -1.524 \\ \hline
\end{tabular}
\caption{Parameter values for the toy model described in Section~\ref{sec:closedbox}. $f_\text{gas}$ and [O/H]$_\text{gas}$ are the median values measured in each radial bin. [O/H]$_\text{final}$ is the best fit to the simple closed-box model given in Equation \ref{eqn:S(Z)}, while [O/H]$_\text{initial}$ is the pre-enrichment found to provide the best fit when using the pre-enriched closed-box model given in Equation \ref{eqn:S(Z)_new}. We used $Z \propto 10^{\text{[O/H]}}$.}
\label{tbl:CDF}
\end{table}

\section{Discussion} \label{sec:discussion}

\subsection{The gas-phase abundance profile}

The panels in Figure~\ref{fig:yeff} provide a clear narrative of star formation and chemical enrichment in $z=2$ star-forming galaxies. The short gas consumption timescale in the center of galaxies means that these regions are more efficient at forming stars. Gas is consumed more rapidly and is depleted, as evinced by the lower gas fraction in the second panel.
Thus, the gas that is left is highly enriched, as shown in the third panel. In this rather simplistic scenario, \emph{the gas metallicity profile is directly set by the gas fraction profile}, in agreement with the prediction of the closed-box model. Outside factors like outflows or mergers, which are definitely present in the TNG simulations, do not cause a substantial change in the metallicity profile. The final panel, showing that the effective yield is remarkably flat out to $\sim 10$ kpc, confirms our simple interpretation: each radial bin can be considered as an independent closed-box system at a different stage of evolution, which is set by the star formation efficiency at that location.
Both real and simulated galaxies are subject to substantial inflows and outflows of gas, and it is therefore surprising that such a simple model is able to capture the gas-phase metallicity profile. However, while gas flows are able to temporarily increase or decrease the metallicity, any subsequent star formation will rapidly bring the system back to the closed-box prediction, as shown in detail by \citet{2007Dalcanton}.

The importance of gas fraction as the primary driver of gas-phase metallicity has been highlighted by several studies. For example, \citet{2019Torrey} showed that in TNG, the mass-metallicity relation is explained by the simple fact that low-mass galaxies have a higher gas fraction, and therefore a lower gas-phase metallicity compared to massive systems. Other types of simulations, including zoom-in simulations, provide similar pictures \citep[e.g.,][]{2016Ma}. These studies were mostly concerned with the interpretation of galaxy-wide metallicities, rather than radial profiles: in this respect, our approach is more similar to that of \citet{2019Belfiore}, who analyze the observed gas metallicity gradients of local galaxies using a simple model consisting of independent radial bins. They adopt the `bath-tub' model rather than a closed box, but reach similar conclusions: the abundance profile is driven by the radial variation of the star formation efficiency. 

\subsection{The stellar abundance profile}

Stellar populations are more difficult to model because they feature, at each location, a distribution of abundances tracing the entire enrichment history, as opposed to the gas abundance which is a single value and mostly traces the current properties of the system.
Indeed, when applying the closed-box model to the stellar abundance distribution we find inconsistent results. 
First, at each radial bin we find a fraction of stars with higher metallicity than the gas, which is not possible for a true closed-box system. Second, the model prediction systematically overproduces metal-poor stars compared to the TNG data. We were able to improve the agreement by introducing a pre-enrichment that strongly declines with radius. At face value, this could indicate that the gas reaching the galaxy centers has already been enriched by star-forming regions, while the outskirts are fed by more pristine gas. However, this is an overly simplistic interpretation because the effect of inflows, outflows, and formation timescales leave a strong imprint on the stellar abundances and cannot be neglected.

A more realistic model of the chemical history would also need to explain the mildly rising [Mg/Fe] profile that characterizes massive galaxies in TNG (see Figure~\ref{fig:profiles}). Using the Versatile Integrator for Chemical Evolution \citep[VICE;][]{2020Johnson}, we were able to produce a rising [Mg/Fe] profile by changing the accretion history as a function of radius. In particular, the inflow in the central regions need to start earlier and last longer compared to the inflow in the outskirts. However, in order to also reproduce the gas fraction, stellar [Mg/H], and gas-phase [O/H] profiles, we need to assume a declining profile for the pre-enrichment value, and a rising profile for the mass loading factor. A detailed investigation of chemical evolution models is beyond the scope of the present work; however, such preliminary exploration confirms that several physical processes must be considered for a proper understanding of the stellar abundance profiles.

Given the complexity of the processes involved, it is remarkable that the stellar abundance profiles are highly similar when plotted as a function of radius in physical units (i.e., kpc), with a scatter of only $\sim0.1$ dex. Moreover, the scatter among the profiles is \emph{larger} when normalizing the radial coordinate to the effective size of each galaxy, as is commonly done (see Figure~\ref{fig:profiles}). This fact, which has important consequences for galaxy-wide measurements as we discuss below, implies that the abundance profiles do not scale with galaxy sizes but are more or less universal, for massive galaxies at $z=2$. To first approximation, the stellar abundance profile is set by the past gas-phase abundance profile, which also follows a roughly universal law, as shown in Figure~\ref{fig:yeff}. Older stars being less enriched does not substantially change the slope of the stellar abundance profile. We thus conclude that, to the first order, the stellar abundance profiles in TNG are ultimately driven by the gas fraction profiles. Metal removal via outflows can certainly play a role, since the closed-box model is inadequate to describe the detailed abundance distributions of stars; however, this effect is not primarily responsible for the overall decline in stellar [Mg/H] by $\sim 0.8$ dex over 10 kpc. This is in agreement with the results of \citet{2019Torrey}, who found that in TNG, the most massive galaxies are \emph{less} efficient at retaining their metals, which is the opposite behavior of what is commonly believed.

\subsection{Interpreting galaxy-wide abundance measurements}

In addition to exploring the physical processes that determine the stellar abundances of massive galaxies, the main goal of this work is to provide a framework for the interpretation of current and future measurements. Stellar abundance measurements are challenging at high redshift, and are often performed on the spatially integrated spectrum of the whole galaxy. Given the steeply declining abundance profiles we find in TNG, our first conclusion is that assessing the physical size probed by the spectroscopic data is paramount. As we discussed in Section~\ref{subsec:aperture_dependence}, such aperture effects can have a major impact on the scaling relations. If one assumes that the spectra are always extracted from the region within $R_e$, then an anti-correlation between stellar abundance and galaxy size naturally emerges: for small galaxies, $R_e$ probes only the central part of the profile, where the stars are more metal-rich, while for large galaxies $R_e$ is able to probe further into the outskirts of the profile, yielding a lower average abundance. This trend, which we find in the TNG population, is also detected in observations at intermediate redshift (see Appendix~\ref{app:0.7scalings}). Our analysis makes a clear prediction: if one could measure the stellar abundances within a fixed physical aperture (e.g., 1 kpc), this trend would disappear or even flip, as shown in Figure~\ref{fig:aperture_dependence_reff}.

Ideally, our predictions should be tested at high redshift, but current measurements at $z\sim2$ suffer from small sample sizes and large uncertainties. The two galaxies with the most accurate abundance measurements, one from \citet{2016Kriek} and the other from \citet{2020Jafariyazani}, have similar masses and redshifts but their [Mg/H] abundances differ by 0.4 dex. On the other hand, the [Mg/Fe] relative abundances are consistent within the error bars. This surprising result may be explained if the measurements for the two galaxies were taken at different physical apertures. In particular, the galaxy studied by \citet{2020Jafariyazani} is strongly lensed, and the spectroscopic observations are therefore able to probe the very central part of the abundance profile, yielding an elevated [Mg/H] measurement. Even so, this aperture effect does not have a strong impact on the measurement of [Mg/Fe], which is predicted to have a nearly flat profile. This interpretation is supported by the stellar abundance gradients measured by \citet{2020Jafariyazani}, which are the only spatially resolved measurements currently available at high redshift. The observed [Fe/H] declines with a gradient of about $-0.040 \pm 0.028$ dex/kpc, while [Mg/Fe] is approximately flat with radius: both measurements are in broad agreement with the typical profiles found in TNG galaxies.

Further observations at $z\sim2$ are required for a robust test of the TNG predictions. With the advent of JWST, stellar abundance measurements will become more accessible at high redshift, and in some cases the spatial profiles will also be measured with high precision.
Given that TNG produces remarkably tight abundance profiles, it will be easy to test this model. If the TNG predictions are correct, one consequence is that the measured abundance pattern cannot be used to constrain the formation timescale of massive galaxies at $z\sim2$ because the intrinsic differences among galaxies, after correcting for aperture effects, are extremely small. 

\subsection{Caveats}
\label{sec:caveats}

Our results are necessarily tied to the specific physical models implemented in the TNG simulations. For example, the FIRE simulations, which feature a much smaller galaxy sample but higher resolution, obtain very diverse metal gradients at $z=2$ \citep[although mostly at lower stellar masses, see][]{2017Ma}. This key difference is likely due to the burstiness of the FIRE star formation model, and may lead to different scaling relations for the stellar abundances. 
Moreover, the absolute value of the metal abundances measured in TNG are highly dependent on the assumed stellar yields, which are still poorly understood. A preliminary comparison of TNG with the stellar abundances observed in local galaxies suggests that the magnesium yields implemented in the simulation may need to be revised \citep{TNG2}. Nonetheless, relative measurements such as radial trends should be substantially more robust than the absolute scale of the abundance values.

Finally, we caution that a proper comparison with the observations should account for biases introduced by the measurement methods. It has been shown that the procedure employed to derive stellar abundances from the simulations (such as the way stellar particles are combined, and whether mass- or luminosity-weighting is adopted) and from the observations (e.g., the choice of stellar libraries or the method used to fit templates to the observed spectra) have potentially large effects on the stellar abundances \citep{2015Guidi, 2016Guidi, 2022Gebek}. Given that our focus is on the high-redshift universe, where current measurements are scarce, we do not attempt a realistic comparison between simulations and observations. This effort will be warranted, however, once the sample of available measurements is substantially increased by new observations with JWST.

\section{Summary \& Conclusions} \label{sec:conclusion}

Analyzing the population of massive ($\log M_\ast/M_\odot > 10.5$) galaxies at $z=2$ in the TNG simulations, we investigated the relations between stellar abundances and other global properties such as stellar mass and effective size; the radial dependence of stellar abundances; and the physical origin for the emerging relations. Here we summarize our findings:
\begin{itemize}

    \item Stellar abundances ([Mg/H] and [Fe/H]) feature a flat relation with stellar mass and an anti-correlation with effective size. The [Mg/Fe] relative abundance has a slightly positive trend with mass and size. Current observations at $z=2$ are in broad agreement with the TNG predictions, but the small sample size and large uncertainties prevent a meaningful comparison. At lower redshift, these relations become progressively flatter and less tight. Galaxies observed at $z=0.7$ follow trends that are similar to those found in the simulations.
    
    \item Galaxy sizes play an important role in the stellar abundance relations. For example, they drive the difference in [Mg/Fe] between galaxies, while it is commonly assumed that this relative abundance mainly tracks galaxy formation timescales. Moreover, the aperture used in determining the stellar abundance is critical: if the abundances are measured within the central kpc instead of within $R_e$, the trend between [Mg/H] and radius is inverted. 
    
    \item Massive galaxies feature a nearly universal [Mg/H] radial profile, which is steeply declining and does not scale with galaxy sizes. This explains the abundance scaling relations and aperture effects: smaller galaxies appear metal-rich simply because most of the stars within $R_e$ are found in the inner part of the abundance profile; on the other hand, larger galaxies have more stars within $R_e$ that are found in the outer parts of the profile, where the abundances are low.
    
    \item Gas-phase metallicity also follows a nearly universal, steeply declining profile. We develop a toy model in which each radial bin is treated as an independent closed-box system: since the effective yield is roughly constant with radius, we can easily interpret the gas abundance profile as solely determined by the gas fraction profile. In practice, at the center of galaxies the gas density is high and gas can be efficiently transformed into stars. This leaves a low gas fraction, which in a closed-box model gives rise to a high metallicity. In the outskirts, on the other hand, star formation proceeds slowly and leaves large gas fractions, resulting in a low gas metallicity.

    \item The stellar abundance profile is set by the gas abundance profile at the time when most stars were formed. To first approximation, then, the stellar abundance profile is also determined by the gas fraction profile. However, our simple closed-box model fails to reproduce the detailed distribution of stellar abundances at a fixed radius. We are able to improve the model by allowing for a radially varying pre-enrichment, but we conclude that a substantially more complex model must be developed in order to explain the stellar abundances in detail.
    
\end{itemize}

Perhaps the most important result from our analysis is the discovery of a nearly universal profile for both stellar and gas abundances in simulated massive galaxies at $z=2$. This sheds light on the physical processes that drive the chemical enrichment of galaxies. At the most fundamental level is the star formation efficiency, which in the TNG model is simply determined by the gas density. This is what ultimately drives the abundance profile: the central regions are faster at transforming gas into stars and therefore their chemical enrichment is accelerated, resulting in higher levels of gas and stellar abundance. Other processes, such as outflows, likely play some role in determining the detailed distribution of chemical abundances, but are not responsible for the steeply declining abundance profiles.

The existence of a nearly universal profile also has important implications for the interpretation of galaxy-wide abundance measurements. In particular, we showed that the stellar abundances are strongly dependent on the size of the aperture used for the measurement. At the same time, this effect provides a robust test for the simulations, because it produces a relatively tight relation between the stellar abundance measured within $R_e$ and the galaxy effective size.

Our conclusions are, naturally, only valid as long as the physical model employed by the TNG simulations is valid.
The extent to which our results apply to the real universe can only be assessed through a comparison with observations.
Since there are intervening processes at lower redshift, such as quenching and galaxy mergers, that tend to dilute the chemical profiles, observations at high redshift are required. The near-infrared capabilities and exquisite sensitivity of JWST will soon make this type of measurements possible for representative samples of high-redshift galaxies, leading to conclusive tests of the theoretical predictions presented in this work. 

\section*{Acknowledgements}

We thank the anonymous referee for their constructive comments. We acknowledge Charlie Conroy and Drew Newman for useful discussions, and Aliza Beverage for providing the abundance measurements. This work started as a project of the Smithsonian Astrophysical Observatory REU program, which is funded in part by the National Science Foundation REU and Department of Defense ASSURE programs under NSF Grants no.\ AST 1852268 and 2050813, and by the Smithsonian Institution. RW is supported by the Natural Sciences and Engineering Research Council of Canada (NSERC), funding reference CITA 490888-16. SB is supported by the Italian Ministry for Universities and Research through the \emph{Rita Levi Montalcini} program.

%%%%%%%%%%%%%%%%%%%%%%%%%%%%%%%%%%%%%%%%%%%%%%%%%%
\section*{Data Availability}

The data underlying this article were accessed from the IllustrisTNG simulation website accessible under https://www.tng-project.org/data/. The derived data generated in this research will be shared on reasonable request to the corresponding author.



\bibliographystyle{mnras}
\bibliography{references} % if your bibtex file is called example.bib

%\begin{thebibliography}{99}
%\end{thebibliography}

%%%%%%%%%%%%%%%%%%%%%%%%%%%%%%%%%%%%%%%%%%%%%%%%%%

%%%%%%%%%%%%%%%%% APPENDICES %%%%%%%%%%%%%%%%%%%%%

\appendix

\section{Stellar Abundances at $\lowercase{z}=0.7$}
\label{app:0.7scalings}

\begin{figure*}
    \centering
    \includegraphics{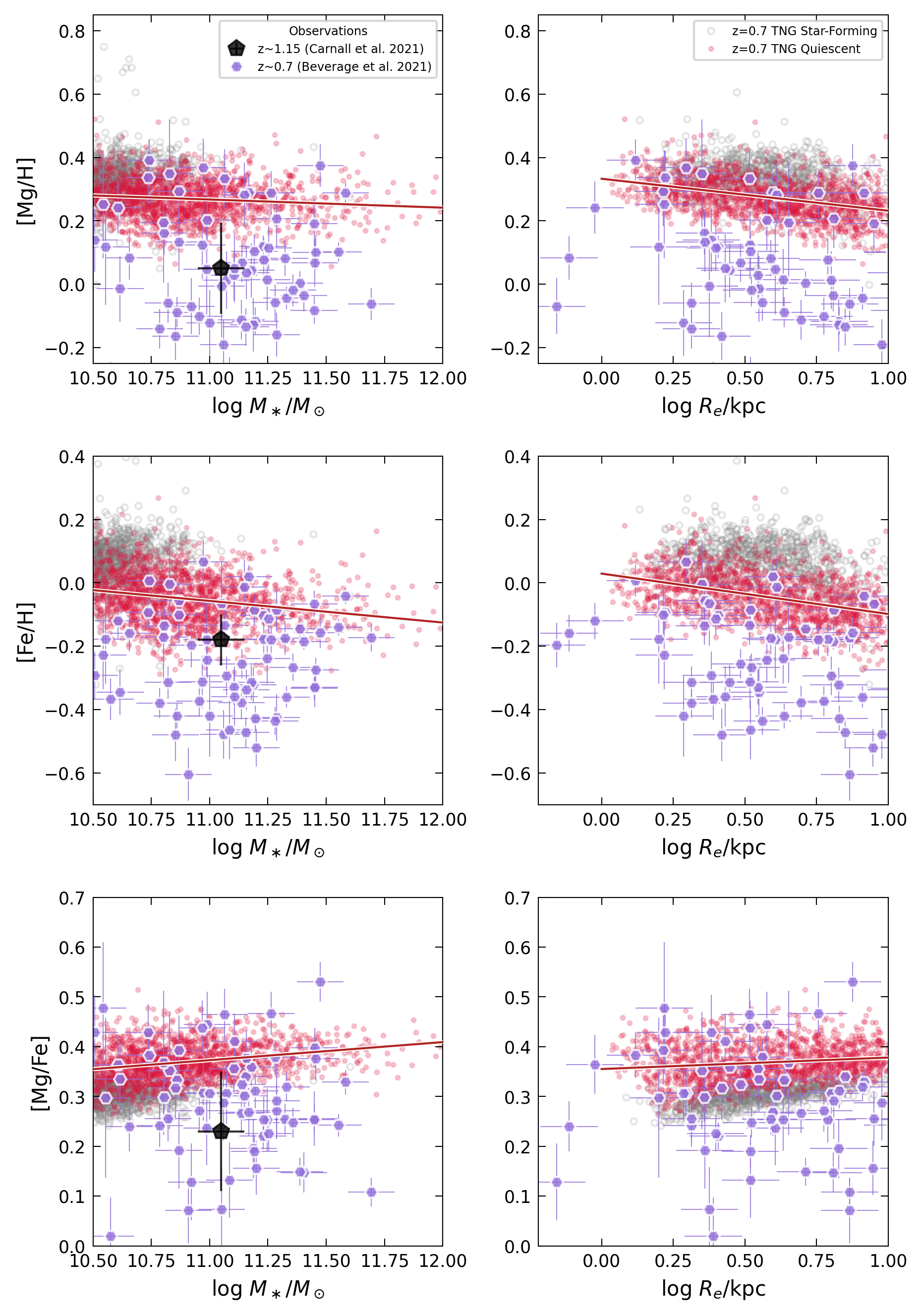}
    \caption{Stellar abundance ratios for galaxies at intermediate redshift as a function of mass (left) and effective radius (right). TNG galaxies at $z=0.7$ are shown as small circles, with quiescent systems in red and star-forming galaxies in gray. In each panel, the red line is a linear fit to the simulated quiescent galaxies. Larger symbols mark observed galaxies \citep{2021Beverage,2021Carnall}}
    \label{fig:z0p7_bigcheddar}
\end{figure*}

In this appendix we explore the stellar abundances of TNG galaxies at $z=0.7$.
Figure~\ref{fig:z0p7_bigcheddar} shows the [Mg/H], [Fe/H], and [Mg/Fe] abundances as a function of stellar mass (left) and effective size (right). The trends are qualitatively similar to those shown in Figure~\ref{fig:z2_bigcheddar} for the $z=2$ population, except for a  negative slope in the relation between [Fe/H] and stellar mass (at $z=2$ this relation is approximately flat, although its slope is formally negative as well). As discussed in Section~\ref{sec:redshift_evolution}, the slope of these relations becomes progressively shallower with cosmic time, and the scatter increases.
In the figure, we also show the measured abundances for observed galaxies. The black marker is from a stack of dozens of galaxies at $z \sim 1.15$ observed by \citet{2021Carnall}, and does not have a meaningful measurement for the effective size. The purple points are individual galaxies observed at $0.6 < z < 0.75$ by the LEGA-C survey \citep{2016vanderWel}, for which \citet{2021Beverage} measured stellar abundances. The higher quality of spectroscopic data that can be obtained for galaxies at $z=0.7$ compared to observations at higher redshifts enables a more detailed comparison between simulations and observations. Generally, we find that measured abundances appear to be 0.2--0.3 dex lower than the simulated values. However, such direct comparison does not take into account several biases, as discussed in Section~\ref{sec:caveats}; moreover, the most recent LEGA-C data release \citep{2021vanderWel} revealed a potential issue with the background subtraction in earlier data releases that may cause the published measured abundances to be systematically underestimated by 0.1--0.2 dex (A. Beverage, private communication). What is most striking in Figure~\ref{fig:z0p7_bigcheddar} is the fact that simulations and observations follow very similar trends. Most notably, Both [Mg/H] and [Fe/H] decline with $R_e$ in the \citet{2021Beverage} sample, with a slope that is fully consistent with the TNG result. This finding supports the existence of a nearly universal abundance profile in real galaxies, which would naturally explain the observed trends.

\begin{figure*}
    \centering
    \includegraphics[width=0.99\textwidth]{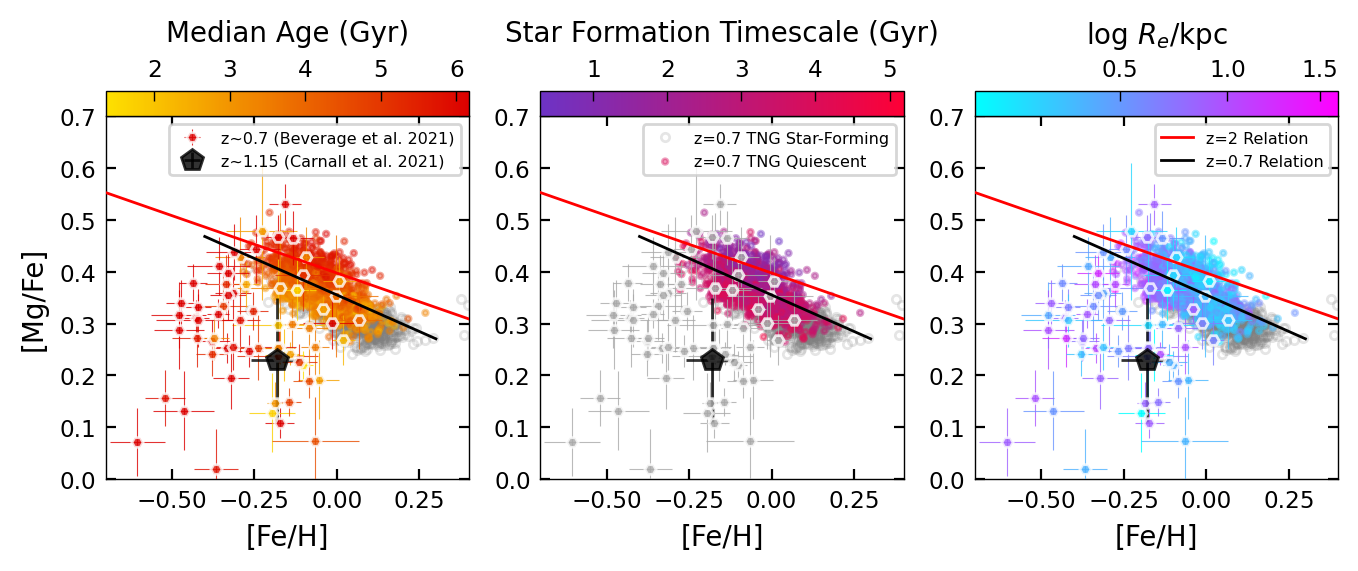}
    \caption{[Mg/Fe] vs [Fe/H] for $z = 0.7$ simulated galaxies. In each panel, quiescent galaxies are color-coded according to a different variable: the median stellar age, the star formation timescale (both calculated within one effective size), and the effective size. 
    We also show the observations at $z\sim0.7$ from \citet{2021Beverage} (color coded according to the measured stellar age and effective size) and \citet{2021Carnall}. The red line is a linear fit to the quiescent simulated galaxies at $z=2$, shown in Figure \ref{fig:z2_three}. The black line is a linear fit to the simulated $z=0.7$ quiescent galaxies.}
    \label{fig:z0p7_three_fullcolor}
\end{figure*}

The same observations are compared to the TNG results on the [Mg/Fe] versus [Fe/H] plane in Figure~\ref{fig:z0p7_three_fullcolor}. For an easy comparison, we include the quiescent $z=2$ trend line in each panel. Here, too, an overall offset towards lower abundances is present in the observations. The scatter is large, and no sequence is evident in the data. The \citet{2021Beverage} galaxies have also been color-coded according to their measured stellar ages (first panel) and effective sizes (last panel); no measurements of the formation timescale are available. The measured stellar ages do not follow the same trend apparent in the TNG simulations; the size trend, on the other hand, is consistent with the TNG prediction, where larger galaxies tend to have lower [Fe/H] but higher [Mg/Fe]. We caution, however, that subtle biases in the observations due to the requirement of high signal-to-noise ratio on the spectra, which generally favors compact and young systems, may affect the comparison with the simulations.

%%%%%%%%%%%%%%%%%%%%%%%%%%%%%%%%%%%%%%%%%%%%%%%%%%

% Don't change these lines
\bsp	% typesetting comment
\label{lastpage}
\end{document}